\documentclass[prl,twocolumn,showpacs,preprintnumbers,aps,nofootinbib]{revtex4-1}
\usepackage{graphicx}
\usepackage{dcolumn}
\usepackage{bm}
\usepackage{amsmath,amssymb,amsfonts}
\usepackage{latexsym}
\usepackage{bbm}
\usepackage{color}
\usepackage{epsfig}

\begin{document}

\preprint{LA-UR-19-30968,
CTPU-PTC-19-28}

\title{
Cancellation mechanism for the electron electric dipole moment
connected with the baryon asymmetry of the Universe}

\author{Kaori Fuyuto$^1$}
\email{kfuyuto@lanl.gov}
\author{Wei-Shu Hou$^2$}
\email{wshou@phys.ntu.edu.tw}
\author{Eibun Senaha$^{3,4,5}$}
\email{eibun.senaha@tdtu.edu.vn}
\affiliation{$^1$Theoretical Division, Los Alamos National Laboratory, Los Alamos, New Mexico 87545, USA}
\affiliation{$^2$Department of Physics, National Taiwan University, Taipei 10617, Taiwan}
\affiliation{$^3$Theoretical Particle Physics and Cosmology Research Group, Advanced Institute of Materials Science, Ton Duc Thang University, Ho Chi Minh City 700000, Vietnam}
\affiliation{$^4$Faculty of Applied Sciences, Ton Duc Thang University, Ho Chi Minh City 700000, Vietnam}
\affiliation{$^5$Center for Theoretical Physics of the Universe, Institute for Basic Science (IBS), Daejeon 34126, Korea}
\bigskip

\date{\today}

\begin{abstract}
We elucidate a cancellation mechanism for the electric dipole moment of
the electron in the general two Higgs doublet model.
The impressive improvement by the ACME Collaboration in 2018
suggests the presence of a new electron Yukawa coupling
that brings in exquisite cancellations among dangerous diagrams,
broadening the solution space for electroweak baryogenesis
driven by an extra top Yukawa coupling.
The cancellation mechanism calls for the new Yukawa couplings to have
hierarchical structures that echo the observed pattern of the
Standard Model Yukawa couplings.
\end{abstract}

\pacs{
}

\maketitle

\paragraph{Introduction.---}
It is remarkable that the Cabibbo-Kobayashi-Maskawa (CKM) framework
is able to explain all laboratory-based measurements of
charge-parity, or $CP$, violation (CPV).
But it is well known that the CPV phase arising from the CKM matrix is
by far insufficient in generating the baryon asymmetry of the Universe (BAU);
hence, some new CPV phase(s) must exist to address this cosmological problem.
Thus, in many well-motivated models beyond the Standard Model (SM),
the existence of such beyond CKM phases is often a common theme.
Detecting the effect of such new CPV phases would provide a powerful probe
of new energy thresholds above the electroweak (EW) scale.

Owing to its high testability, EW baryogenesis (EWBG)~\cite{ewbg}
is of primary importance and broad interest.
However, data from the Large Hadron Collider, such as the
measurement of Higgs boson properties, have diminished or even completely
eliminated the EWBG parameter space in most models.
Complementary to
 collider probes,
extreme low-energy searches such as the electric dipole moment (EDM)
 of the electron, neutron, etc., have put further stress on models.
In particular, with the new upper bound on the electron EDM (eEDM),
$|d_e|<1.1\times 10^{-29}\;e\,\text{cm}$ at 90\% Confidence level (C.L.),
given by the ACME collaboration in 2018 (ACME18)~\cite{Andreev:2018ayy},
many EWBG scenarios are now in jeopardy.
Although calculations of CPV sources still have significant
 uncertainties,
hence the amount of BAU might go upward by more refined analyses,
the impact of the ACME18 bound is nevertheless overwhelming.
In other words, if EWBG is the true mechanism behind BAU,
the unprecedented ACME18 result may indicate
some undisclosed mechanism that renders $d_e$ small.

In a previous paper~\cite{Fuyuto:2017ewj},
we have explored the general two Higgs doublet model (g2HDM),
i.e. without the {\it ad hoc} discrete $Z_2$ symmetry,
in which an additional $3\times 3$ Yukawa coupling matrix
for each type of charged fermion should be $CP$ and flavor violating.
It was shown that the extra top Yukawa couplings, naturally
$\mathcal{O}$(1) in magnitude, can provide sufficient CPV needed for BAU.
The specific scenario exemplified in Ref.~\cite{Fuyuto:2017ewj} is now excluded by ACME18,
but one should explore more generic parameter space
to see how one can survive the ACME18 bound.

\begin{figure}[b]
\center
\includegraphics[width=3.8cm]{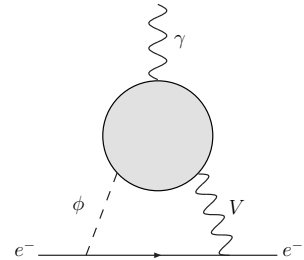}
\caption{
Two-loop Barr-Zee diagrams~\cite{Barr:1990vd} contributing to the electron EDM,
where $\phi$ denotes neutral and charged Higgs bosons,
and $V$ denotes vector bosons $\gamma$, $Z$ and $W$.}
\label{de_BZ}
\end{figure}

In this paper, we find a built-in cancellation mechanism
among the diagrams of Fig.~\ref{de_BZ} that
can evade the ACME18 bound, and support EWBG via top transport in g2HDM.
The new bound suggests the existence of a new electron Yukawa coupling that,
in conjunction with the extra top Yukawa coupling,
can render eEDM sufficiently small.
The cancellation mechanism works only when
the hierarchical structure of the new Yukawa couplings is
close to those of the SM Yukawa couplings
and with particular pattern of CPV phases,
which may reflect an underlying flavor structure in g2HDM.

\paragraph{g2HDM, EWBG, and ThO EDM.---}
The g2HDM extends SM by adding one extra Higgs doublet~\cite{Branco:2011iw},
 but without imposing a $Z_2$ symmetry.
With 
flavor changing neutral Higgs couplings controlled by~\cite{Hou:2017hiw} 
fermion mass and mixing hierarchies
 {\it plus} alignment
 [i.e., rather close proximity of $h(125)$ to the SM Higgs boson], the
phenomenological consequences of the g2HDM is much richer than usual 2HDMs
with $Z_2$ symmetries~\cite{Branco:2011iw}.

The Yukawa interactions in the mass eigenbasis are
\begin{align}
- \mathcal{L}_Y = \bar{f}y_{\phi}^f R f\phi
+ \bar{f}_\uparrow \Big[V\rho^{f_\downarrow}R
                       - \rho^{f_{\uparrow}\dagger}V L \Big]f_\downarrow H^+
+ \text{H.c},
\label{Yuk}
\end{align}
where $f=u,d,e$; $f_\uparrow=u,\nu$; $f_\downarrow=d, e$;
${L,R} = (1\mp \gamma_5)/2$;
$\phi = h, H, A$ are the neutral scalars and $H^+$ is the charged scalar;
and $V$ is the CKM matrix for quarks and unit matrix for leptons.
In Eq.~(\ref{Yuk}), $\rho^f$ are $3\times 3$ Yukawa matrices
which are new sources of $CP$ and flavor violation, and
$y_\phi^f$ are related $3\times 3$ matrices with elements
\begin{align}
y_{hij}^f &= \frac{\lambda_i^f}{\sqrt{2}}\delta_{ij}s_{\gamma}+\frac{\rho_{ij}^f}{\sqrt{2}}c_{\gamma},
 \label{hff}\\
y_{Hij}^f&=\frac{\lambda_i^f}{\sqrt{2}}\delta_{ij}c_{\gamma}-\frac{\rho_{ij}^f}{\sqrt{2}}s_{\gamma}, \\
y_{Aij}^{f_{\uparrow}}&= -i\frac{\rho_{ij}^{f_\uparrow}}{\sqrt{2}},\quad
y_{Aij}^{f_{\downarrow}}= i\frac{\rho_{ij}^{f_\downarrow}}{\sqrt{2}}
\label{yAij}
\end{align}
where $\lambda_i^f=\sqrt{2}m_i^f/v~(v=246~\text{GeV})$,
$s_{\gamma}=\sin\gamma$, $c_{\gamma}=\cos\gamma$,
and alignment implies $\cos^2\gamma$ is quite small.
{We will comment later on the further mixing between
 $h,H$, and $A$ induced by CPV phases of $\rho_{ij}^f$ at one-loop level.

As far as EWBG is concerned, not all complex phases are relevant.
As found in Ref.~\cite{Fuyuto:2017ewj},
$|\rho_{tt}|\gtrsim 0.01$ with moderate CPV phase can generate sufficient BAU, while
${\cal O}(1)$ $\rho_{tc}$ with maximal phase
can also play a role in case $|\rho_{tt}|\lesssim 0.01$.
Even though the $\rho_{tt}$ mechanism is more efficient,
the parameter space is severely constrained by ACME18.
In the $\rho_{tc}$ mechanism, on the other hand, in exchange for
less efficient baryogenesis, it does not induce dangerous eEDMs by itself.
The two mechanisms are therefore complementary.
In this work, we focus exclusively on the $\rho_{tt}$ case
and parametrize $\rho_{ij}=|\rho_{ij}|e^{i\phi_{ij}}$.

The effective EDM for thorium monoxide (ThO) is given by~\cite{Chupp:2014gka, Chupp:2017rkp}
\begin{align}
d_{\text{ThO}}=d_e+\alpha_{\text{ThO}}C_S,
\end{align}
where $d_e$ comes from the dimension-5 operator
$-\frac{i}{2} d_e(\bar{e}\sigma^{\mu\nu}\gamma_5e)F_{\mu\nu}$
 with $F_{\mu\nu}$ the electromagnetic field strength tensor,
 while
the second term arises from nuclear spin-independent
electron-nucleon interaction described by $
-\frac{G_F}{\sqrt{2}}C_S(\bar{N}N)(\bar{e}i\gamma_5 e)$,
 where $G_F$ is the Fermi constant.
ACME18 gives~\cite{Andreev:2018ayy}
$d_{\text{ThO}}=(4.3\pm 4.0)\times 10^{-30}~e~\text{cm}$,
with the stated bound on $d_e$ obtained by assuming $C_S=0$.
With the estimate~\cite{Fuyuto:2018scm} of $\alpha_{\text{ThO}}=1.5\times10^{-20}$,
as we will see below, $C_S$ cannot be completely neglected in our case,
so we shall use $d_{\text{ThO}}$ of ACME18 to explore the model.

In g2HDM, the dominant contributions to $d_e$ come from the Barr-Zee diagrams~\cite{Barr:1990vd}, as depicted in Fig.~\ref{de_BZ}, which
we decompose into three pieces, depending on
the particles attached to the electron line. That is,
\begin{align}
d_e = d_e^{\phi\gamma}+d_e^{\phi Z}+d_e^{\phi W},
\label{de}
\end{align}
where $\phi$ can be the neutral $h,H,$ and $A$ bosons or the $H^+$ boson.
$CP$ is violated at the lower and/or upper vertices of the $\phi$ line.
It is known that $d_e^{\phi\gamma}$ gives the dominant contribution
among the three pieces; hence, the cancellation must occur in this sector.
We note, however, that, although $d_e^{\phi Z}$ and $d_e^{\phi W}$ are subleading,
they are not always smaller than the ACME18 bound.

We further decompose each $d_e^{\phi V}$ in Eq.~(\ref{de}) into
three types of diagrams, consisting of fermions; $W$ and $H^+$ loops
 for $d_e^{\phi\gamma}$ and $d_e^{\phi Z}$;
and $f_\uparrow/f_\downarrow$, $W/\phi$, and $H^\pm/\phi$ loops for $d_e^{\phi W}$.
These are denoted as $(d_e^{\phi V})_i$, $i=f, W, H^+$ for $V=\gamma, Z$,
and $(d_e^{\phi W})_i$, $i=f_\uparrow/f_\downarrow, W/\phi, H^\pm/\phi$.

\begin{figure}[t]
\center
\includegraphics[width=3.7cm]{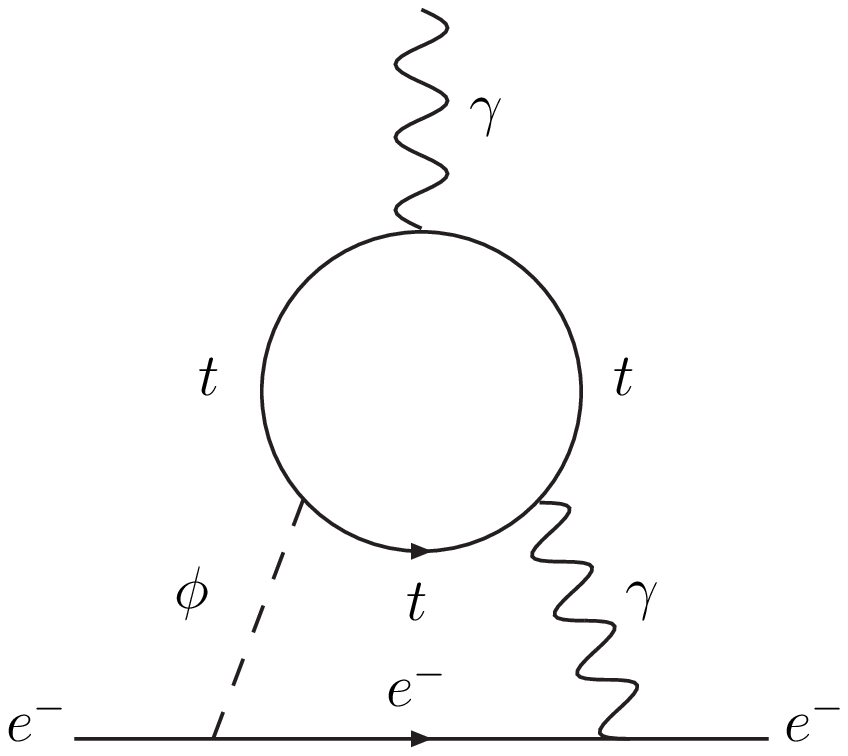}
\hspace{0.2cm}
\includegraphics[width=3.7cm]{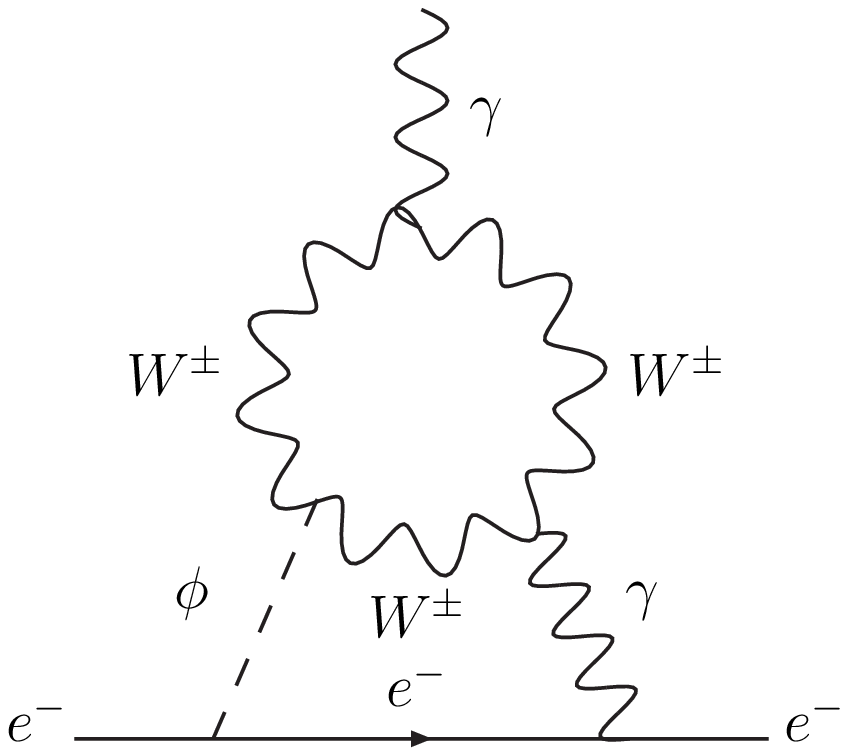}
\vspace{0.2cm}
\caption{
Two dominant diagrams in which $\text{Im}\rho_{ee}$ enters.}
\label{de_Hgam}
\end{figure}

If $\rho_{tt}$ is the only element that has nonzero CPV phase
{and other $\rho$ elements are zero},
one would have $C_S=0$, and $d_e$ hence $d_{\rm ThO}$ is
solely induced by $(d_e^{\phi \gamma})_t$,
which is the left diagram of Fig.~\ref{de_Hgam}.
We find
\begin{align}
\frac{(d_{e}^{\phi\gamma})_{t}}{e}
&= \frac{\alpha_{\rm em}s_{2\gamma}}{12\sqrt{2}\pi^3v}
	\frac{m_e}{m_t}\text{Im}\rho_{tt}\Delta g, \nonumber\\
&= -6.6\times 10^{-29}
\left(\frac{s_{2\gamma}}{0.2}\right)\left(\frac{\text{Im}\rho_{tt}}{-0.1}\right)
\left(\frac{\Delta g}{0.94}\right),
\label{de_Hgam_t_0}
\end{align}
where $e$ is the positron charge,
 $\alpha_{\text{em}}=e^2/4\pi$
and $\Delta g = g(m_t^2/m_h^2)-g(m_t^2/m_H^2)$, and the loop function
$g$ is defined in Ref.~\cite{Barr:1990vd}.

In the second line of Eq.~(\ref{de_Hgam_t_0}),
we take one of the benchmark points considered in Ref.~\cite{Fuyuto:2017ewj},
i.e. $c_\gamma=0.1$, $m_h=125$ GeV and $m_H=m_A=500$ GeV,
which is now excluded by ACME18.
This could be circumvented by making $\text{Im}\rho_{tt}$ and/or $c_\gamma$ small.
For instance, $|(d_{e}^{\phi\gamma})_{t}|$ would become smaller than the ACME18 bound
if $|\text{Im}\rho_{tt}|\lesssim 0.01$, without changing the value of $c_\gamma$.
However, this would no longer be the $\rho_{tt}$-driven EWBG
scenario~\cite{Fuyuto:2017ewj}.
For smaller $c_\gamma$, the dependence of BAU on $c_\gamma$
has not been studied yet in g2HDM.
But
since $c_\gamma\to 0$ corresponds to the SM-like limit,
the variation of the vacuum expectation value ratio $\Delta \beta$ during electroweak phase transition
would be suppressed with decreasing $c_\gamma$.

We conclude that the $\rho_{tt}$-driven EWBG case
as stated above~\cite{Fuyuto:2017ewj} is unlikely to survive the ACME18 bound.

\paragraph{Cancellation mechanism for ThO EDM.---}
In Ref.~\cite{Fuyuto:2017ewj} we set $\rho_{ee} = 0$ for simplicity,
but there is no symmetry or mechanism to make it zero exactly.
Once complex $\rho_{ee}$ comes in, $(d_e^{\phi\gamma})_W$ as
shown in the right diagram of Fig.~\ref{de_Hgam} can be comparable
or even bigger than $(d_e^{\phi \gamma})_t$,
which is analogous to $h$ decay to a diphoton.

To elucidate our cancellation mechanism,
we decompose $(d_{e}^{\phi\gamma})_{i}$ into two parts
\begin{align}
(d_{e}^{\phi\gamma})_i \equiv (d_{e}^{\phi\gamma})_{i}^{\text{mix}}+(d_{e}^{\phi\gamma})_{i}^{\text{extr}},
\end{align}
where the first term arises from the mixing between SM and extra Yukawa couplings,
while the second term is purely from extra Yukawa couplings.
For the top-loop contribution, one has
\begin{align}
\frac{(d_{e}^{\phi\gamma})_{t}^{\text{mix}}}{e}
&= \frac{\alpha_{\rm em}s_{2\gamma}}{12\sqrt{2}\pi^3v}
	\left[\text{Im}\rho_{ee}\Delta f+\frac{m_e}{m_t}\text{Im}\rho_{tt}\Delta g\right],\label{de_Hgam_t} \\
\frac{(d_{e}^{\phi\gamma})_{t}^{\text{extr}}}{e}
&\simeq
\frac{\alpha_{\text{em}}}{12\pi^3m_t}\text{Im}(\rho_{ee}\rho_{tt})\Big[f(\tau_{tA})+g(\tau_{tA})\Big],
\label{de_Hgam_t_ex}
\end{align}
where {$\tau_{ij}=m_i^2/m_j^2$, $\Delta X = X(\tau_{th})-X(\tau_{tH})$,}
with $X=f,g$ defined in Refs.~\cite{Barr:1990vd}
being monotonically increasing loop functions,
so $\Delta X >0$ for $m_h < m_H$.
For $(d_{e}^{H\gamma})_{t}^{\text{extr}}$, we take the approximation of
$c_\gamma \ll 1$ and $m_H\simeq m_A$
in order to see the structure of the cancellation mechanism more clearly. 
In our numerical analysis, however, we do not take any approximations for $(d_{e}^{\phi\gamma})_{t}^{\text{extr}}$
and confirm that this approximation does not spoil the essential point.

For the $W$-loop contribution, on the other hand, there is no extra Yukawa coupling in the $\phi WW$ vertex, so the
$(d_e^{\phi\gamma})_W$ is solely given by $(d_{e}^{\phi\gamma})_W^{\text{mix}}$,
which is
\begin{align}
\frac{(d_{e}^{\phi\gamma})_W^{\text{mix}}}{e}
&=-\frac{\alpha_{\text{em}}s_{2\gamma}}{64\sqrt{2}\pi^3v}\text{Im}\rho_{ee}
\Delta\mathcal{J}^\gamma_W,\label{de_Hgam_W}
\end{align}
where $\Delta\mathcal{J}^\gamma_W=\mathcal{J}^\gamma_W(m_h)-\mathcal{J}^\gamma_W(m_H)$,
with $\mathcal{J}^\gamma_W$ defined in Ref.~\cite{Abe:2013qla},
which is a monotonically decreasing function;
hence, $\Delta\mathcal{J}^\gamma_W > 0$ for $m_h < m_H$.

We consider the cancellation
$(d_e^{\phi\gamma})_t^{\text{mix}} + (d_e^{\phi\gamma})_W^{\text{mix}} = 0$,
under the condition that $(d_e^{\phi\gamma})_t^{\text{extr}} = 0$.
The case of having $(d_e^{\phi\gamma})_t^{\text{extr}} \neq 0$ is discussed later.
From Eqs.~(\ref{de_Hgam_t}), (\ref{de_Hgam_t_ex}) and (\ref{de_Hgam_W}),
these two conditions lead, respectively, to
\begin{align}
\frac{\text{Im}\rho_{ee}}{\text{Im}\rho_{tt}}
& = c\times \frac{\lambda_e}{\lambda_t},\quad
\frac{\text{Re}\rho_{ee}}{\text{Re}\rho_{tt}} = -
\frac{\text{Im}\rho_{ee}}{\text{Im}\rho_{tt}},
\label{de_cancel_phigam}
\end{align}
where $c=(16/3)\Delta g/(\Delta\mathcal{J}_W^\gamma-(16/3)\Delta f)$.
For instance, $c\simeq0.71$ for $m_h=125$ GeV and $m_H=500$ GeV.
Combining the two conditions in Eq.~(\ref{de_cancel_phigam}), one gets
$|\rho_{ee}/\rho_{tt}| = c \times \lambda_e/\lambda_t$,
with a correlated phase between $\rho_{tt}$ and $\rho_{ee}$.
Note that $c$ is not sensitive to the exotic Higgs spectrum
that is consistent with first-order electroweak phase transition and
hence does not change drastically in the parameter range for EWBG.

With the above cancellation, $d_e^{\phi Z}$, $d_e^{\phi W}$ and $C_S$
become potentially important. We estimate~\cite{Dekens:2018bci} $C_S$ as
\begin{align}
C_S &= -2v^2
\bigg[
	6.3\,(C_{ue}+C_{de})+C_{se}\frac{41~\text{MeV}}{m_s}\nonumber\\
&\hspace{1.05cm}
	+C_{ce}\frac{79~\text{MeV}}{m_c} + 0.062
	\left(\frac{C_{be}}{m_b}+\frac{C_{te}}{m_t}\right)
\bigg],
\end{align}
where $C_{qe}$ is defined by $\mathcal{L}_{4f}^{\text{CPV}} = \sum_qC_{qe}(\bar{q}q)(\bar{e}i\gamma_5 e)$,
which emerges after integrating out all neutral Higgs bosons.
The quark mass suppressions are canceled
  by corresponding Yukawa couplings in $C_{qe}$,
  so all quark flavors are generically relevant.
Note that for $s_\gamma\simeq 1$ and $m_H\simeq m_A$, $C_{qe}$
for $u$- and $d$-type quarks are cast in the form of
$C_{ue} \simeq \text{Im}(\rho_{ee}\rho_{uu})/(2m_A^2)$ and
$C_{de} \simeq \text{Im}(\rho_{ee}\rho_{dd}^*)/(2m_A^2)$, respectively,
which implies that $C_{qe}\simeq 0$ if $(d_e^{\phi\gamma})_{q}^{\text{extr}}\simeq 0$.

Before turning to numerical results, we comment on CPV effects at one-loop level,
where $h$ and $H$ can mix with $A$ through ${\rm Im}\rho_{tt}$ and ${\rm Im}\rho_{ee}$ 
and hence are no longer $CP$ eigenstates.
The mass eigenstates are obtained by $(H_1,H_2, H_3)^T=O \, (h, H, A)^T$,
where $O$ is an orthogonal matrix that
diagonalizes the Higgs mass squared matrix $\mathcal{M}_N^2$, i.e.,
$O^T\mathcal{M}^2_NO = \text{diag}(m_{H_1}^2, m_{H_2}^2,m_{H_3}^2)$.
The dominant contributions to the $CP$-mixing entries are
$(\mathcal{M}^2_N)_{13} = -3\lambda_t\,\text{Im}\rho_{tt}\,m_t^2/4\pi^2$
and $(\mathcal{M}^2_N)_{23} = -3\,\text{Re}\rho_{tt}\,\text{Im}\rho_{tt}\,m_t^2/4\pi^2$.
For $\phi_{tt}=-90^\circ$, one finds that
$\theta_{13} \simeq \tan^{-1}\big[2(\mathcal{M}^2_N)_{13}/(m_h^2-m_A^2)\big]/2
 \simeq 9.6\times 10^{-3}$ for $|\rho_{tt}|=1$ and $m_A=500$ GeV,
and the effects are small enough to be ignored.
For $\phi_{tt}\neq -90^\circ$, on the other hand,
despite $(\mathcal{M}^2_N)_{23}$ being loop induced,
the 2-3 mixing angle would be
$\theta_{23} \simeq \tan^{-1}\big[2(\mathcal{M}^2_N)_{23}/(m_H^2-m_A^2)\big]/2
 \simeq 45^\circ$ if $m_H\simeq m_A$,
and $H$ and $A$ cannot be identified as $CP$ eigenstates at all.
But even for this case, $d_e$ would not be much affected
because of the orthogonality of the matrix $O$.
For example, we estimate the relevant part for $(d_e^{\phi\gamma})_t$
as $\sum_i O_{2i}O_{3i}f(m_t^2/m_{H_i}^2) \simeq
O_{21}O_{31}f(m_t^2/m_{H_1}^2) + (O_{22}O_{32}+O_{23}O_{33})f(m_t^2/m_{H_2}^2)\ll 1$,
where $m_{H_2} \simeq m_{H_3}$ and $\sum_iO_{2i}O_{3i}=0$ have been used.
We conclude that the one-loop CPV effects are rather minor.

\begin{figure*}[t]
\center
\includegraphics[width=7.7cm]{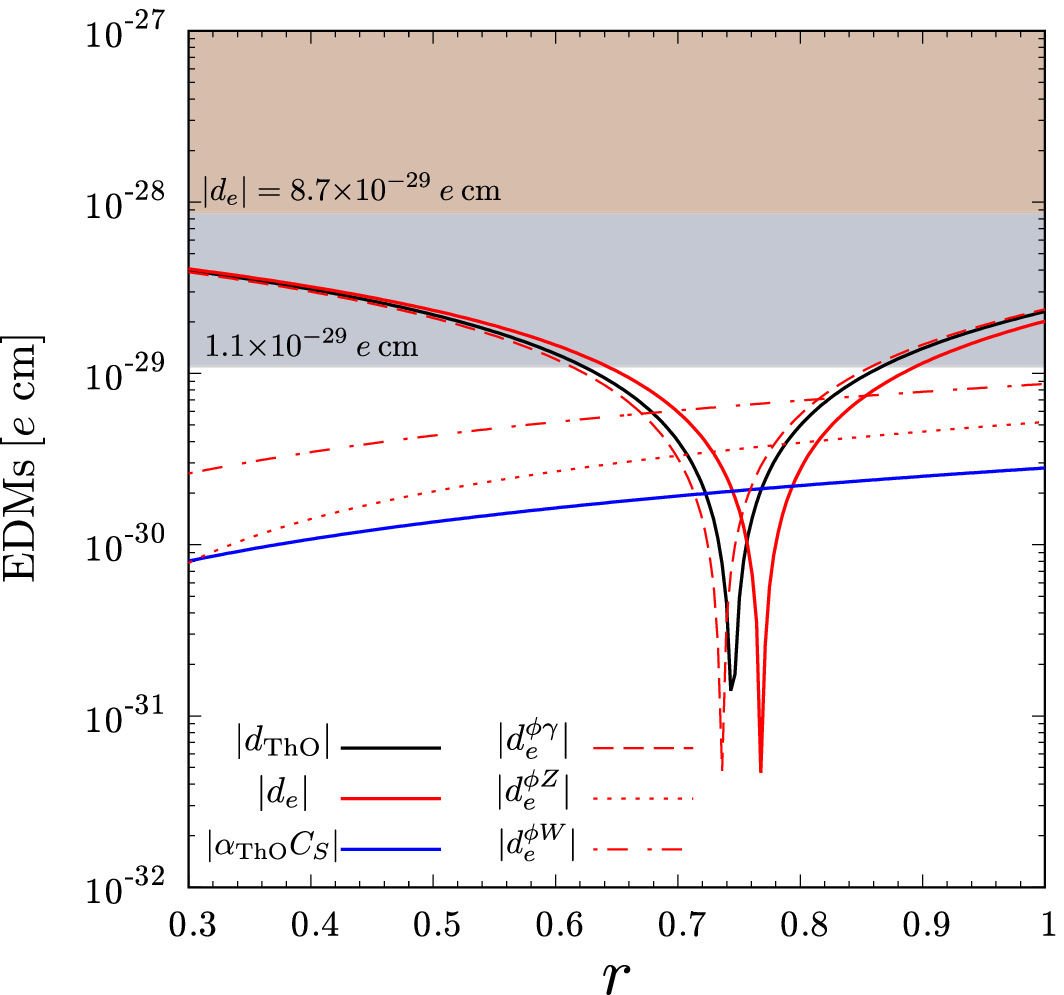}
\hspace{0.5cm}
\includegraphics[width=7.3cm]{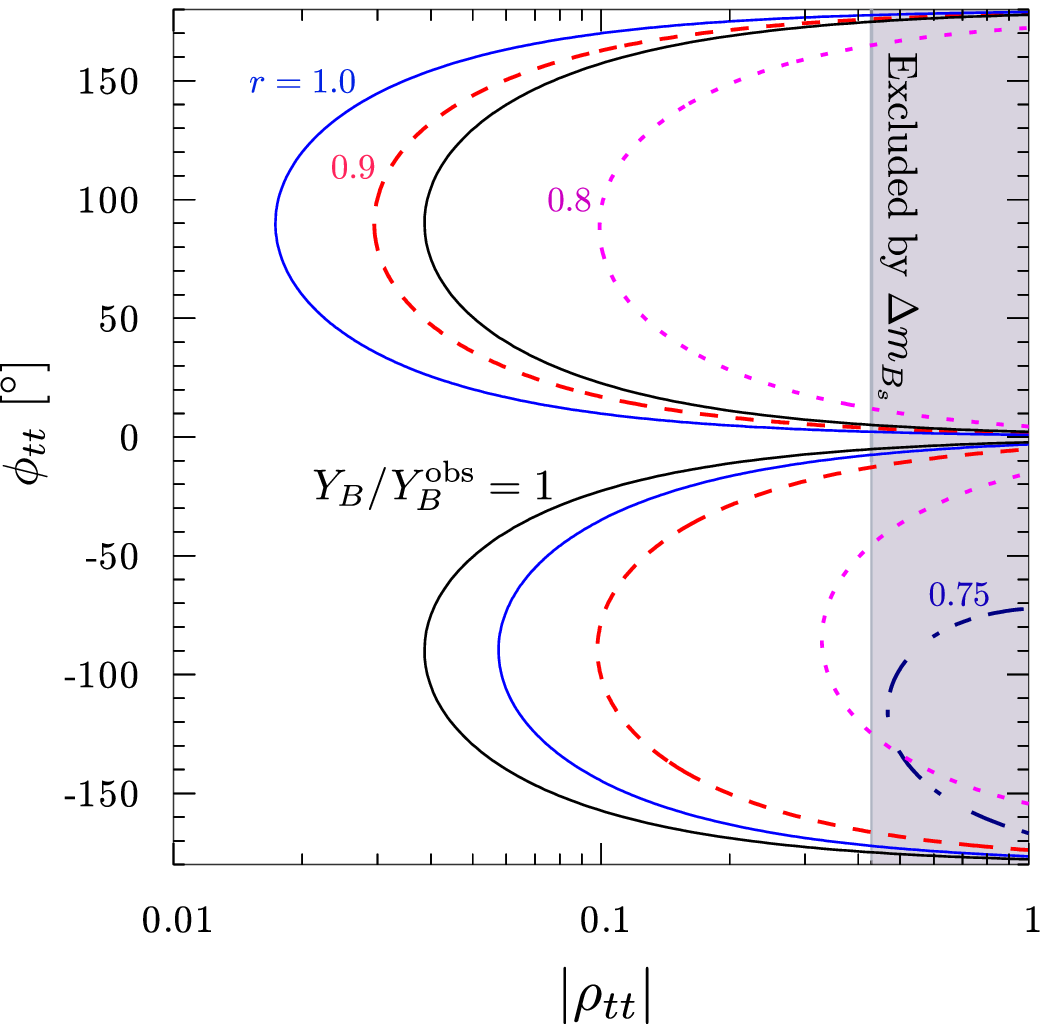}
\caption{
(Left) $|d_{\rm ThO}|$ and its details as functions of $r$,
where $\text{Re}\rho_{ff}=-r\lambda_f\text{Re}\rho_{tt}/\lambda_t$ and
$\text{Im}\rho_{ff}=r\lambda_f\text{Im}\rho_{tt}/\lambda_t$
for charged fermion $f$.
We have taken $\text{Re}\rho_{tt}=\text{Im}\rho_{tt}=-0.1$,
as well as $c_\gamma=0.1$ and $m_H=m_A=m_{H^\pm}=500$ GeV.
The bounds from ACME are overlaid.
(Right) The 2$\sigma$-allowed region of $d_{\text{ThO}}$ with $r=1.0$ (blue, solid), 0.9 (red, dashed), 0.8 (magenta, dotted) and 0.75 (navy blue, dot-dashed).
The region to the right of the black solid contour, $Y_B/Y_B^{\text{obs}}=1$,
 is allowed,
while the gray shaded region is excluded by $B_s$-$\bar{B}_s$ mixing.
Other input parameters are the same as in the left plot.}
\label{dThO}
\end{figure*}

\paragraph{Numerical results.---}

We choose $\rho_{tt}$ to be consistent with successful EWBG (for details, see Ref.~\cite{Fuyuto:2017ewj})
and parametrize the other diagonal $\rho_{ff}$ elements as
$\text{Re}\rho_{ff}=a_f(\lambda_f/\lambda_t)\text{Re}\rho_{tt}$ and
$\text{Im}\rho_{ff}=b_f(\lambda_f/\lambda_t)\text{Im}\rho_{tt}$,
where $a_f$ and $b_f$ are real parameters such that $|a_f|=|b_f|\equiv r_f$.
From the argument given above, the cancellation mechanism would be at work
if $a_e<0$ and $b_e>0$.
In what follows, we consider a flavor-blind scaling of $a_f=-r$ and $b_f=r$.

To see the cancellation behavior,
we first investigate the magnitude of $d_{\text{ThO}}$.
In Fig.~\ref{dThO} (left) we plot $|d_{\text{ThO}}|$ (black, solid) and
its compositions $|d_e|$ (red, solid), $|\alpha_{\text{ThO}}C_S|$ (blue, solid), $|d_e^{\phi\gamma}|$ (red, dashed),
$|d_e^{\phi Z}|$ (red, dotted), $|d_e^{\phi W}|$ (red, dot-dashed) as functions of $r$,
where we set $\text{Re}\rho_{tt}=\text{Im}\rho_{tt}=-0.1$ as an illustrative point of successful EWBG, as we explicitly show below.
The ACME18~\cite{Andreev:2018ayy} and previous~\cite{Baron:2013eja} (ACME14) bounds
are shown as the
{gray and brown} shaded regions as marked.
The absence of $\rho_{ee}$ would correspond to the case of $r=0$,
with $d_e\simeq (d_e^{\phi\gamma})_t$ estimated in Eq.~(\ref{de_Hgam_t_0}).
This specific point~\cite{Fuyuto:2017ewj} is excluded by ACME18.
The situation changes considerably, however, for $r\neq0$.

As can be seen, strong cancellation occurs in $d_e^{\phi\gamma}$
around $r \simeq 0.75$. This is owing to the presence of $(d_e^{\phi\gamma})_W$, and $d_e^{\phi W}$ becomes dominant, followed by $d_e^{\phi Z}$,
shifting the cancellation point in $d_e$ upward.
However, the dip in $d_{\text{ThO}}$ moves downward due to the $C_S$ contribution.
In any case, $d_{\text{ThO}}$ can be suppressed, characteristically, by two orders of magnitude below
the ACME18 bound
owing to the cancellation mechanism.\footnote{Even though we could have $d_{\text{ThO}}\simeq0$ by finely tuning $r$, the precise value of $r$ would not be meaningful since it is subject to high-order corrections that are missing here.}

We display, in Fig.~\ref{dThO} (right), the 2$\sigma$ allowed region
of $d_{\text{ThO}}$ in the ($|\rho_{tt}|$, $\phi_{tt}$) plane,
taking $r=1.0$ (blue, solid), 0.9 (red, dashed), 0.8 (magenta, dotted)
 and 0.75 (navy blue, dot-dashed), respectively.
The region to the left of these contours is allowed,
while to the right of the black contours corresponds to
$Y_B>Y_B^{\text{obs}}=8.59\times 10^{-11}$~\cite{Ade:2013zuv} for EWBG~\cite{Fuyuto:2017ewj}.
The gray shaded region for larger $|\rho_{tt}|$ values
is excluded by
$B_s$-$\bar{B}_s$ mixing~\cite{Amhis:2016xyh}.
Note that in Ref.~\cite{Fuyuto:2017ewj},
we considered $\phi_{tt} < 0$ for BAU positive.
However, one can have $\phi_{tt} > 0$ by flipping the sign of $\Delta\beta$.
Since the central value of $d_{\text{ThO}}$ is positive,
the allowed region is asymmetric in $\phi_{tt}$.
For $r=1.0$ and 0.9, only $\phi_{tt} < 0$ is consistent with $\rho_{tt}$-driven EWBG,
but $\phi_{tt} > 0$ becomes possible as $r$ approaches
the cancellation point at $r \sim 0.75$, enlarging the room for $\rho_{tt}$-driven EWBG.

Let us comment on the case in which $(d_e^{\phi\gamma})_t^{\text{extr}} \neq 0$.
Taking $\text{Re}\rho_{ee}\simeq  0$ for illustration, we find
\begin{align}
\frac{\text{Im}\rho_{ee}}{\text{Im}\rho_{tt}} \simeq \frac{(16/3)\Delta g}{\Delta\mathcal{J}_W^\gamma-(16/3)\Delta f+\epsilon}\frac{\lambda_e}{\lambda_t}\equiv c' \times\frac{\lambda_e}{\lambda_t},
\end{align}
where $s_{2\gamma}\lambda_t \, \epsilon = -(16/3)\text{Re}\rho_{tt}\big[ f(\tau_{tA})+g(\tau_{tA})\big]$.
Thus, the coefficient $c$ in Eq.~(\ref{de_cancel_phigam}) can be altered by $\epsilon$,
where $|c'|$ can become much larger than 1
when $\epsilon$ makes the denominator small.
But then $d_{\text{ThO}}$ gets too large due to sizable $\text{Im}\rho_{ee}$ and is
hence inconsistent with ACME18.
We find $|c'|\gtrsim 0.3$ for experimentally allowed $\text{Re}\rho_{tt}$, 
so the cancellation mechanism still suggests that
the $\rho$ matrices follow the SM Yukawa coupling hierarchy.
It is also worth mentioning that, despite the small parameter space,
further cancellation in $d_{\text{ThO}}$ can occur if we
take flavor-dependent $a_f$ and $b_f$ such that $|a_f|$, $|b_f|<1$.
In this case, $\rho_{bb}$ could play an elevated role.

We remark that rephasing-invariant CPV quantities~\cite{Botella:1994cs} 
involving two Yukawa couplings are proportional to 
$\sum_i\lambda_i\text{Im}\rho_{ii}$, with $i$ the generation index.
Therefore, any relationships among $\text{Im}\rho_{ff}$ should be associated with the SM Yukawa couplings
(the most significant being $\lambda_t\, {\rm Im}\rho_{tt}$ for EWBG),
which gives an intuitive understanding of our cancellation mechanism.

Some comparisons with previous work is in order.
Ref.~\cite{Bian:2014zka} discusses EDM cancellations in 2HDMs with softly broken $Z_2$.
Since only one physical CPV phase exists in such models, 
cancellation is caused by a specific choice of $\tan\beta$ 
rather than a new electron Yukawa coupling.
Ref.~\cite{Egana-Ugrinovic:2018fpy} does consider 
a cancellation mechanism in 2HDM without $Z_2$, 
but the analysis assumes $\text{Re}\rho_{ff}=0$, and cancellations with $\text{Re}\rho_{ff} \neq 0$ are not studied.
Furthermore, the extra Higgs bosons are taken as decoupled, and
hence there is no EWBG connection.

Before closing, we note that the ACME14 bound was confirmed by an independent experiment
using the polar molecule ${}^{180}\text{Hf}{}^{19}\text{F}^+$~\cite{Cairncross:2017fip}.
Given the significance of the ACME18 result, it should be similarly cross-checked,
preferably using different methods.
It is quite interesting that,
while the largest diagonal extra Yukawa coupling,
$\rho_{tt}$, is responsible for BAU, it works in concert
with the smallest diagonal extra Yukawa coupling,
$\rho_{ee}$, to generate an eEDM that might be revealed
soon by very-low-energy, ultraprecision probes.
We look forward to updates on electron EDM that may further probe
the parameter space of $\rho_{tt}$-driven EWBG.

\paragraph{Conclusion.---}

In the scenario in which an extra Yukawa coupling $\rho_{tt}$ drives EWBG,
we demonstrate that the ACME18 result suggests the presence of
a new electron Yukawa coupling, bringing in an exquisite cancellation mechanism
for eEDM measured in ThO, which broadens the parameter space.
This cancellation can be at work only when the hierarchical structure of
the new Yukawa couplings is similar to those of the SM Yukawa couplings,
which may reflect some underlying flavor structure in the general 2HDM.
Alternatively, EWBG may be due to the weaker mechanism from
flavor changing $\rho_{tc}$ coupling that evades the eEDM bound.

\vskip0.2cm
\noindent{\bf Acknowledgments} \
We thank Jordy de Vries for helpful discussions.
K.F. is supported by
the LANL/LDRD Program,
E.S. is supported in part by IBS under the project code IBS-R018-D1,
and W.S.H. is supported by grants MOST 106-2112-M-002-015-MY3, and
NTU 108L104019, 108L893301.


\end{document}